\pdfoutput=1
\documentclass[
aip,
 amsmath,amssymb,
 reprint,%
]{revtex4-1}

\usepackage{graphicx}% Include figure files
\usepackage{dcolumn}% Align table columns on decimal point
\usepackage{bm}% bold math
\usepackage[utf8]{inputenc}
\usepackage[T1]{fontenc}
\usepackage{mathptmx}
\usepackage{bm}
\usepackage{amsmath}
\usepackage{commath}
\usepackage{upgreek}
\usepackage{graphicx}
\usepackage{epstopdf}
\usepackage{xcolor}

\begin{document}

%\preprint{}

\title{The nonlinear dynamo effect of tearing modes} %Title of paper



\author{Luo Yuhang}
\email[]{luo-yh15@mails.tsinghua.edu.cn}
\author{Gao Zhe}
%\thanks{}
%\altaffiliation{Department of Engneering Physics, Tsinghua University, Beijing, 100084}
\affiliation{Department of Engineering Physics, Tsinghua University, Beijing, 100084}


%\date{\today}
\newcommand{\Abs}[1]{\vert {\bm{#1}}\rvert}
\newcommand{\mAbs}[1]{\vert \overline{\bm{#1}}\rvert}
\newcommand{\AbsSqr}[1]{\vert \bm{#1}\rvert^{2}}
\newcommand{\mAbsSqr}[1]{\vert \overline{\bm{#1}}\rvert^{2}}
\newcommand{\fAbsSqr}[1]{\vert \widetilde{\bm{#1}}\rvert^{2}}
\newcommand{\fvct}[1]{\widetilde{\bm{#1}}}
\newcommand{\mvct}[1]{\overline{\bm{#1}}}
\newcommand{\mBfour}{\vert\overline{\bm{B}}\rvert^{4}}
\newcommand{\Sqr}[1]{\vert#1\rvert^{2}}

\begin{abstract}
The nonlinear dynamo effect of tearing modes is derived with the resistive MHD equations. 
The dynamo effect is divided into two parts, parallel and perpendicular to the magnetic field. 
Firstly, the force-free plasma is considered. 
It is found that the parallel dynamo effect drives opposite current densities at the different sides of the rational surface, making the $\lambda =\bm{j}\cdot\bm{B}/\AbsSqr{B}$ profile completely flattened near the rational surface.
There are many rational surfaces for the turbulent plasma, which means the plasma is tending to relax into the Taylor state.
In contrast, a bit far from the rational surface, the parallel dynamo effect is much smaller, and the nonlinear dynamo form approximates the quasilinear form.
Secondly, the pressure gradient is included. 
It is found that rather than the $\lambda$ profile, the $\bm{j}\cdot\bm{B}$ profile is flattened by the parallel dynamo effect. 
Besides, the perpendicular dynamo effect of tearing modes is found to eliminate the pressure gradient near the rational surface.
% current density perpendicular to $\bm{B}$.
In addition, our result also provides another basis for the assumption that current density is flat in the magnetic island for the tearing modes theory. 
\end{abstract}

\pacs{}% insert suggested PACS numbers in braces on next line

\maketitle %\maketitle must follow title, authors, abstract and \pacs

\section{Introduction}
%\label{}
%\subsection{}
%\subsubsection{}
The dynamo effect\cite{Moffatt1984, Sokoloff2014} is an average electric motive force produced by velocity and magnetic field fluctuations. 
In astrophysical plasmas, the dynamo effect is widely discussed as a mechanism of magnetic flux amplification.\cite{Yoshizawa1990, Brandenburg2001}
In fusion plasmas, the dynamo effect has been invoked in explaining the relaxation process of self-organized systems like Reversed-Field Pinch (RFP)\cite{Hokin1991} and Spheromak\cite{Al-Karkhy1993, Jarboe2012}.

A typical relaxation process in RFP and Spheromak contains four stages. 
Firstly, the current density profile is modified by the external current drive method. 
Generally, the driven current concentrates at the edge of plasma when using non-inductive procedures like helicity injection and wave injection. 
Secondly, the modified current density triggers instabilities, which are mainly considered as tearing modes. 
Thirdly, the overlap of the growing instabilities leads to turbulent plasma. 
Fourthly, the turbulent plasma relaxes into a state with a stable current density profile. 
The first three stages are intuitive, while the last stage is hard to understand.

At the fourth stage, Taylor\cite{Taylor1986} shows that plasma will relax into the state where the magnetic energy is minimized while the magnetic helicity is conserved relative to the magnetic energy. 
This state is called the Taylor state. 
According to Taylor's theory, $\bm{j}=\lambda\bm{B}$ holds in the relaxation area, and the parameter, $\lambda =\bm{j}\cdot\bm{B}/\AbsSqr{B}$, is constant. 
This means the pressure gradient goes to zero; thus, the confinement is not well. 
In reality, the plasma would be stable before fully relaxing into the Taylor state, which makes plasma with a finite pressure gradient can be obtained. 
Taylor specifies to what state the turbulent plasma will relax, but how this relaxation occurs remains unsolved. 

The dynamo effect has been widely discussed these years to explain the relaxation process's detail, both theoretically\cite{Strauss1985, Bhattacharjee1986, Ji2002} and numerically\cite{Cappello2006, Mirnov2004, Bonfiglio2005}.
A lot of experiments\cite{Ji1996, DenHartog1999, Fontana2000, Brower2002} have been developed to observe and measure the dynamo effect.

In the most widely studied MHD dynamo model, the dynamo effect is expressed as $\langle \fvct{v}\times \fvct{B} \rangle$,  $ \fvct{v}$ and  $\fvct{B}$ correspond to the fluctuating velocity and magnetic field respectively, $\langle~\rangle$ indicates average over poloidal and toroidal direction. 

The dynamo effect can be divided into two components. 
One component is parallel to the mean magnetic field, $\varepsilon_\parallel=\langle \fvct{v}\times \fvct{B} \rangle \cdot \mvct{B}/ \mAbs{B}$, which is also called the $\alpha$ effect\cite{Seehafer1996,Ji2002}. 
This component is important in driving parallel current.\cite{Ji1995, Brower2002} 
The other component is perpendicular to the mean magnetic field and radial direction, $\varepsilon_\perp=\langle \fvct{v}\times \fvct{B} \rangle \cdot (\mvct{B}\times \nabla r)/ \mAbs{B}$, which plays a crucial role in driving perpendicular current density and change the pressure gradient. 

The parallel dynamo effect induced by the tearing fluctuations is first obtained by Strauss\cite{Strauss1985}. 
The turbulent plasma is decomposed into many tearing modes in Strauss's theory. 
Each tearing mode is considered separately, which means the interactions of the tearing modes are ignored here.
Bhattacharjee et al. \cite{Bhattacharjee1986} also found a similar parallel dynamo effect form of the tearing modes. 
The velocity and magnetic fluctuations are derived using linear tearing equations in their theories. 
Therefore the form of the parallel dynamo effect is quasilinear.
The quasilinear form of parallel dynamo effect clearly shows a flattening effect of $\lambda$ profile. 
This makes the theory popular in the explanation of Taylor's relaxation process.

However, the quasilinear approach is only suitable in the linear growth stage of the tearing modes. 
When the tearing modes evaluate the nonlinear growth stage and the final saturated state, the quasilinear theory is not applicable. 
To our best knowledge, the nonlinear theory of the parallel dynamo effect has not been developed yet. 

The perpendicular dynamo effect is omitted in Strauss\cite{Strauss1985}'s theory since the force-free plasma is considered. 
Bhattacharjee et al. \cite{Bhattacharjee1986} used the interchange mode to give the nonlinear form of the perpendicular dynamo effect.
The tearing mode induced perpendicular dynamo effect is also not discussed yet.

The tearing mode has been widely discussed.
The linear theory of a single tearing mode is first derived by Furth et al. \cite{Furth1963}. 
Rutherford\cite{Rutherford1973} proposed the nonlinear theory in which the quasilinear modified current was considered. 
The convincing theory of multiple tearing modes interactions has not been developed. 
So that we also treat the turbulent plasma as multiple individual tearing modes in this paper. 
Furthermore, the nonlinear treatment of a single tearing mode is used to give the nonlinear form of the dynamo effect.

Our treatment of the tearing mode is a little different from most tearing mode theories did.
We do not perform the flux average procedure, which is extensive used in most tearing mode theories. 
As we mentioned earlier, the plasma we study here is turbulent. 
Thus the fast transport along with the magnetic flux no longer exists. 

This paper gives the nonlinear result of both parallel and perpendicular dynamo effect in the nonlinear tearing mode theory frame without using flux average. 
We find the driven current by the nonlinear parallel dynamo effect in the force-free plasma totally flattens $\lambda$ near the rational surface. 
In terms of the turbulent plasma, the dynamo effect should be significant because there are many rational surfaces.
This makes the $\lambda$ gradient vanished in the turbulent area, which means relaxing into the Taylor state. 

When the distance goes far from the rational surface, the parallel dynamo effect becomes smaller.
Our nonlinear form approximates well with the quasilinear form as the distance goes really far from the rational surface.
From a quantitative view, we can ignore the parallel dynamo effect in the region far from the rational surface. 
This makes the quasilinear form of the dynamo effect less critical in explaining the relaxation process.

Furthermore, the apparent singularity in the parallel dynamo effect's quasilinear form makes the quasilinear theory not suitable near the rational surface of the tearing mode. 
Although Strauss declared that the inclusion of inertia would remove the singularity, we will demonstrate that as soon as performing the nonlinear tearing mode procedure, the singularity is no longer exists when using the MHD model.

When the pressure gradient is included, we find the perpendicular dynamo effect will drive an opposite current density to cancel the mean perpendicular current density.
We also find the nonlinear form of the parallel dynamo effect is different from that in the force-free plasma.
The parallel dynamo effect flattens $\bm{j}\cdot\bm{B}$ instead of $\lambda$.
The constant $\lambda$ is required by the Taylor state in the force-free plasma as mentioned above.
The requirement of $\bm{j}\cdot\bm{B}$ may give an idea of extending Taylor's theory into the plasma with the pressure gradient.

Because the tearing modes' interactions are not concerned in our theory, our result is also suitable for a single tearing mode.
For most single tearing mode theories, the flattened current density is a premise hypothesis when performing flux average.
Our result gives another perspective of how the current density flattens near the rational surface for a single tearing mode.

This paper is organized as follows. The resistive MHD model is presented in Sec. \ref{sec:model}. The force-free plasma is considered in Sec. \ref{sec:forcefree}, where the nonlinear form of parallel dynamo effect is obtained and compared with the quasilinear form. In Sec. \ref{sec:pressure}, the nonlinear forms of both the parallel and perpendicular dynamo effect are derived with the pressure gradient existed. Finally, the conclusions and discussion are included in Sec. \ref{sec:conclusion}.

\section{\label{sec:model} Physical Model}
This section presents the definition of the dynamo effect under the frame of the resistive MHD model.
Consider the full resistive MHD equations:
\begin{align}
&\rho\left(\frac{\partial \bm v}{\partial t} + \bm v \cdot \nabla \bm v\right) = \bm j \times \bm B - \nabla p \label{mmteqn},\\
&\bm E + \bm v \times \bm B = \eta \bm j \label{ohmeqn},\\
 &\frac{\partial \bm B}{\partial t}=-\nabla\times\bm E \label{frdeqn},\\
 &\bm j =  \nabla \times \bm B\label{ampeqn}.
\end{align}\par
Gaussian units is used here, light speed $c=1$ and $4\pi$ is absorbed for convenience. 
Equation~(\ref{mmteqn}) is the momentum or force balance equation where $\rho$ is the mass density, $\bm v$ is the plasma velocity, $\bm j$ is the current density, and $p$ is the pressure. 
Eq.~(\ref{ohmeqn}) is the ohm's law where $\bm E$ is the electric field, and $\eta$ is the resistivity which is assumed to be constant. 
Eq.~(\ref{frdeqn}) is Faraday's law, and Eq.~(\ref{ampeqn}) is Ampère's circuital law, and the displacement current is neglected.\par

The cylindrical coordinate system$(r,\theta,\phi)$ is employed, $r$, $\theta$ are the radial and poloidal direction, respectively, and $\phi=z/R$ indicates the toroidal direction with $R$ the major radius and $z$ the axial direction. 
The large aspect ratio limit is not implied in this paper, making our theory more suitable for small aspect radio systems. At the same time, toroidal effects are not considered here.\par

All the physical quantities are separeted into mean and fluctuating parts, for example, the magnetic field
\begin{align}
&\bm B=\mvct{B}(r,t) + \fvct{B}(r,\theta,\phi,t),\\
 &\mvct{B}  = \frac{1}{4\uppi^2}\oint \bm B \dif\theta \dif\phi.
\end{align}
The perturbation part $\fvct{B}$ takes the form of  $\mathrm{e}^{\mathrm{i}(m \theta- n \phi)}$ for a tearing mode with the polodial number $m$ and torodial number $n$.
The mean magnetic field satisfies the resonant condition $\bm k\cdot \mvct{B}=0$ at rational surface $r=r_\mathrm{s}$ , where $\bm k=m\nabla\theta-n\nabla\phi$ is the constructed mode number vector.\par

In the following derivation, we introduce a vector potential $\bm A$ and an electrostatic potential $U$.
 Here the Coulomb gauge $\nabla \cdot \bm{A} = 0$ is used. Therefore, 
\begin{align}
&\bm B = \nabla \times \bm A,\\
&\bm E = -\nabla U - \frac{\partial \bm A}{\partial t}.
\end{align} \par
The Ohm's law Eq.~(\ref{ohmeqn}) becomes
\begin{equation}\label{mohmeqn}
\frac{\partial \bm A}{\partial t} =\bm v \times \bm B -\nabla U -  \eta \bm{j}.
\end{equation}\par
If we averages  Eq.~(\ref{mohmeqn}) over $\theta$ and $\phi$ and remain the $\langle \fvct{v}\times \fvct{B} \rangle$ term, the dynamo induced current density can be expressed as 
\begin{equation}
 \delta \mvct{j} =\frac{1}{\eta} \left< \fvct{v}\times \fvct{B} \right>.
\end{equation}\par
To calculate the form of dynamo effect, the velocity perturbation, $\fvct{v}$, should be represented as a function of magnetic perturbation, $\tilde{\bm B}$. 
Take linearization of Eq.~(\ref{mohmeqn}) and cross with $\mvct{B}/\mAbsSqr{B}$, we get
\begin{equation}\label{veqn}
\fvct{v}_\perp = -\nabla \tilde{U}\times \frac{\mvct{B}}{\mAbsSqr{B}}-\frac{\eta \fvct{j} \times \mvct{B}}{\mAbsSqr{B}}-\frac{\partial\fvct{A}}{\partial t}\times \frac{\mvct{B}}{\mAbsSqr{B}}.
\end{equation}\par
The parallel dynamo effect is 
\begin{equation}\begin{split}\label{dfeqn}
\varepsilon_\parallel &=\left< \fvct{v}\times \fvct{B} \right>\cdot\frac{\mvct{B}}{\mAbs{B}}=\\
&\qquad\qquad-\frac{\fvct{B}_\perp^*\cdot\nabla \tilde{U}}{2\mAbs{B}}-\frac{\fvct{B}_\perp^*}{2\mAbs{B}}\cdot\left(\eta\fvct{j}+\frac{\partial \fvct{A}}{\partial t}\right).
\end{split}\end{equation}

The perpendicular dynamo effect is 
\begin{equation}\label{dfpeqn}
\varepsilon_\perp=\left< \fvct{v}\times \fvct{B} \right>\cdot\frac{\mvct{B}\times\nabla r}{\mAbs{B}}=\frac{\tilde{B}_r^*\fvct{v}\cdot\mvct{B}-\fvct{B}^*\cdot\mvct{B}\tilde{\upsilon}_r}{2\mAbs{B}}.
\end{equation}

The perpendicular dynamo effect does not occur in the force-free plasma because the current density is always parallel to the magnetic field, and the perpendicular current can not be driven.

Dynamo effect equals zero in the perfect conducting plasma because of the ``frozen-in'' restriction. 
The resistivity must be included in the analysis of the dynamo effect.

\section{\label{sec:forcefree} the force-free plasma}
In this section, we will give the nonlinear form of the dynamo effect in the force-free plasma. 
Both the steady state and the growth stage are considered. 
We will also discuss the difference between the quasilinear form and our nonlinear form.

In force-free plasma, $\bm{j}\times\bm{B}=0$ always holds, and all the pressure terms are neglected. 
Taking  $\bm B/\AbsSqr{B}$ component of Eq.~(\ref{mohmeqn}) to eliminate $\bm v$, we get the relation of $\tilde U$ and $\tilde \lambda$
\begin{equation}\label{ldeqn}
\eta \tilde \lambda = -\frac{\mvct{B}}{\mAbsSqr{B}}\cdot\nabla \tilde U  - \left(\frac{\partial\bm A}{\partial t}\cdot \frac{\bm B}{\AbsSqr{B}}\right)_1.
\end{equation}
where $()_1$ indicates the fluctuating part, and
\begin{equation}
\tilde \lambda = \left(\frac{\bm{j} \cdot \bm{B}}{\AbsSqr{B}}\right)_1.
\end{equation}\par
Crossing Eq.~(\ref{mmteqn}) with $\bm B/\AbsSqr{B}$, taking divergence and then linearizing gives 
\begin{equation}\label{mmmteqn}
\nabla \cdot \left(\frac{\overline\rho}{\mAbsSqr{B}}\frac{\partial \fvct{v}}{\partial t}\times \mvct{B}\right)=\fvct{B}\cdot\nabla\overline\lambda+\mvct{B}\cdot\nabla\tilde\lambda,
\end{equation}
where $\overline\lambda = \mvct{j}\cdot \mvct{B}/\mAbsSqr{B}$ is the mean part of $\lambda$, and $\nabla \cdot \bm j=0$ is used.\par
Substituting Eq.~(\ref{ldeqn}) and linearized Eq.~(\ref{mohmeqn}) into Eq.~(\ref{mmmteqn}) leads to
\begin{equation}\begin{split}\label{lineqn}
&\frac{\partial}{\partial t}\nabla \cdot \left(\frac{\overline\rho}{\mAbsSqr{B}}\nabla \tilde{U}\right)-\frac{(\bm k\cdot \mvct{B})^2 }{\eta \mAbsSqr{B}}\tilde U=\\
&\qquad\qquad\qquad\fvct{B}\cdot\nabla\overline\lambda-\frac{\mathrm{i}\bm k\cdot \mvct{B}}{\eta}\left(\frac{\partial \bm A}{\partial t}\cdot \frac{\bm B}{\AbsSqr{B}}\right)_1.
\end{split}\end{equation}
The time derivatives of $\mvct{B}$ and $\overline{\rho}$ are assumed to be neglected, and $\overline\rho/\mAbsSqr{B}$ is assumed constant. 
%This procedure is the same as Strauss\cite{Strauss1985} did. 
Eq.~(\ref{lineqn}) is obtained using linear theory, and the nonlinear term will be added in the following discussion.\par
We replace $\overline\lambda$ by $\overline\lambda+\delta\overline\lambda$ in Eq.~(\ref{lineqn}), where 
\begin{equation}\label{lmdeqn}
\delta\overline{\lambda} = \frac{\delta \mvct{j}\cdot \mvct{B}}{\mAbsSqr{B}}=-\frac{\fvct{B}_\perp^*\cdot\nabla \tilde{U}}{2\eta\mAbsSqr{B}}-\frac{\fvct{B}_\perp^*}{2\eta\mAbsSqr{B}}\cdot\left(\eta\fvct{j}+\frac{\partial \fvct{A}}{\partial t}\right).
\end{equation}
This idea is first proposed by Rutherford\cite{Rutherford1973}. The adding of the quasilinear term $\delta\overline\lambda$ into the linear theory Eq.~(\ref{lineqn}) leads to the nonlinear theory equation,
\begin{equation}\begin{split}\label{nlineqn}
&\frac{\partial}{\partial t}\nabla \cdot \left(\frac{\overline\rho}{\mAbsSqr{B}}\nabla \tilde{U}\right)-\frac{(\bm k\cdot \mvct{B})^2 }{\eta \mAbsSqr{B}}\tilde U=\\
&\qquad\qquad\fvct{B}\cdot\nabla\left(\overline\lambda+\delta\overline\lambda\right)-\frac{\mathrm{i}\bm k\cdot \mvct{B}}{\eta}\left(\frac{\partial \bm A}{\partial t}\cdot \frac{\bm B}{\AbsSqr{B}}\right)_1.
\end{split}\end{equation}
Note that this is the equation govern the plasma behavior in the resistive layer. 
Usually, most of the tearing mode theories perform the asymptotic matching of the inner resistive and the outer ideal region to get the growth rate. 
However, we will just focus on the resistive layer. 
Calculate $\tilde{U}$ from Eq.~(\ref{nlineqn}) in terms of other quantities, then obtain the form of dynamo effect using Eq.~(\ref{dfeqn}). 

Using Eqs.~(\ref{dfeqn}) and (\ref{lmdeqn}) to substitute $\delta\overline\lambda$ in Eq.~(\ref{nlineqn}) gives
\begin{widetext}
\begin{equation}\begin{split}\label{ffnlineqn}
\frac{\partial}{\partial t}\nabla \cdot \left(\frac{\overline\rho\nabla \tilde{U}}{\mAbsSqr{B}}\right)+\fvct{B}\cdot\nabla\left(\frac{\fvct{B}_\perp^*}{2\eta\mAbsSqr{B}}\cdot\frac{\partial \fvct{A}}{\partial t}\right)+\frac{\mathrm{i}\bm k\cdot \mvct{B}}{\eta}\left(\frac{\partial \bm A}{\partial t}\cdot \frac{\bm B}{\AbsSqr{B}}\right)_1=\fvct{B}\cdot\nabla\overline\lambda- \fvct{B}\cdot\nabla\frac{\fvct{B}_\perp^*\cdot(\nabla \tilde{U}+\eta\fvct{j})}{2\eta\mAbsSqr{B}}+\frac{(\bm k\cdot \mvct{B})^2 }{\eta \mAbsSqr{B}}\tilde U.
\end{split}\end{equation}
\end{widetext}
All the time derivative terms are on the left hand side of the equation.
The $\overline\lambda$ gradient term is reserved, because the flux average assumption can not be applied for the turbulent plasma.
Li\cite{Li1995} did not use flux average either in his tearing mode theory, but he did not include this $\lambda$ gradient term in his model. 
So that our solutions of the velocity and magnetic field fluctuations are different from those Li obtained.

The quasilinear theories\cite{Strauss1985, Bhattacharjee1986} did not include the nonlinear term, $\fvct{B}\cdot\nabla\delta\overline\lambda$, which is of great importance in the nonlinear growth stage and steady state of a tearing mode. 
As we can see, $\delta\overline\lambda$ is actually driven by the parallel dynamo effect.

\subsection{Steady state}
Firstly, the steady state is considered, which means the time derivative terms are omitted.
Eq.~(\ref{ffnlineqn}) becomes
\begin{equation}\label{ssffnlineqn}
 \fvct{B}\cdot\nabla\frac{\fvct{B}_\perp^*\cdot\nabla \tilde{U}}{2\eta\mAbsSqr{B}}-\frac{(\bm k\cdot \mvct{B})^2 }{\eta \mAbsSqr{B}}\tilde U=\fvct{B}\cdot\nabla\overline\lambda-\fvct{B}\cdot\nabla\frac{\fvct{B}_\perp^*\cdot\fvct{j}}{2\mAbsSqr{B}}.
\end{equation}

Some simplifications will be made in order to analyse Eq.~(\ref{ssffnlineqn}). $\mAbsSqr{B}=\text{constant}$, because $p+\mAbsSqr{B}/2=\text{constant}$ in the cylindrical geometry and $\nabla p$ is neglected in a force-free plasma. 
The ``constant-$\psi$'' approximation\cite{Furth1963} is also applied as the resistive layer is assumed narrow, limiting the treatment only suitable for $m\geq 2$ modes. 
Therefore, $\tilde{B}_r$ is assumed constant and the another component of $\fvct{B}_\perp$, $\mvct{B}\times\nabla r \cdot \fvct{B}_\perp$, is assumed zero.

Taylor expansion of $\overline\lambda$ around the rational surface gives 
\begin{equation}\label{eqn:ld}
\overline\lambda = \overline\lambda_0+\frac{\partial \overline\lambda}{\partial s}\Big|_{s=0}s+\cdots, \quad\text{where}~ s=r-r_s.
\end{equation}
The form of $\bm k\cdot \mvct{B}$ is determined by $\overline\lambda$, and the leading order is 
\begin{equation}\label{kBapprox}
\bm k\cdot \mvct{B}\approx |\bm{k}||\mvct{B}|\overline\lambda_0 s,
\end{equation} where the orders greater than $\mathcal{O}(s)$ are neglected.\par
The $\fvct{A}$ is assumed containing two components,
\begin{equation}
\fvct   {A} = \frac{\fvct{A}\cdot\mvct{B}}{\mAbsSqr{B}}\mvct{B}+\tilde{A}_r\nabla r.
\end{equation}
The $\mvct{B}$ component of $\fvct{A}$ can be expressed as $\tilde{B}_r$, and we only keep the leading term,
\begin{equation}\label{ABapprox}
\frac{\fvct{A}\cdot\mvct{B}}{\mAbsSqr{B}}=\frac{\tilde{B}_r}{\mathrm{i}\bm{k}\times\mvct{B}\cdot\nabla r}\approx \frac{\tilde{B}_r}{\mathrm{i}|\bm{k}||\mvct{B}|} .
\end{equation}\par
And using Coulomb gauge, $\nabla \cdot \fvct{A}=0$, $\tilde{A}_r$ can be expressed as
 \begin{equation}\label{Areqn}
\frac{\partial \tilde{A}_r}{\partial r}=-\mathrm{i} \bm{k}\cdot\mvct{B}\frac{\fvct{A}\cdot\mvct{B}}{\mAbsSqr{B}}\approx -\overline\lambda_0\tilde{B}_r s,
\end{equation}
which makes 
\begin{equation}\label{jreqn}
\tilde{j}_r=\nabla^2\tilde{A}_r\approx -\overline\lambda_0\tilde{B}_r.
\end{equation}
Therefore, the second term of Eq.~(\ref{ssffnlineqn})'s right hand side(RHS) becomes
\begin{equation}\label{fBpjeqn}
\fvct{B}\cdot\nabla\frac{\fvct{B}_\perp^*\cdot\fvct{j}}{2\mAbsSqr{B}}=\fvct{B}\cdot\nabla\frac{\overline\lambda_0\Sqr{\tilde{B}_r}}{2\mAbsSqr{B}}
\end{equation}
At most fusion plasma, the fluctuating magnetic field is much smaller than mean magnetic field, so this term can be neglected when compared to the first term of Eq.~(\ref{ssffnlineqn})'s RHS.

Comparison of the two terms of Eq.~(\ref{ssffnlineqn})'s left hand side(LHS) shows that
\begin{equation}\label{eqn:radio}
 \fvct{B}\cdot\nabla\biggl(\frac{\fvct{B}_\perp^*\cdot\nabla \tilde{U}}{2\eta\mAbsSqr{B}}\biggr)\bigg/\frac{(\bm k\cdot \mvct{B})^2 }{\eta \mAbsSqr{B}}\tilde U \approx \frac{\vert\tilde{B}_r\rvert^2}{2(\bm{k}\cdot\mvct{B})^2\tilde{U}}\frac{\partial^2 \tilde{U}}{\partial r^2}.
\end{equation}

Although, the structure of magnetic island doesn't exist in the turbulent plasma, we also use the ``island width'', $w$, as a characteristic length of the magnetic perturbation, which can be written as
\begin{equation}
w=4\left(\frac{|\tilde{B}_r|}{|\bm{k}||\mvct{B}|\overline\lambda_0}\right)^{1/2}.
\end{equation}

The radial derivative can be written as radial characteristic length $L$. 
The electrostatic potential fluctuation, $\tilde{U}$, is corresponding to fluctuating velocity field from Eq.~(\ref{veqn}). 
Therefore, $L$ is also the space scale of velocity fluctuation.

From the flow structure of nonlinear tearing mode\cite{Rutherford1973}, we can assume $L\approx w/4$. Therefore, the ratio, Eq.~(\ref{eqn:radio}), becomes
\begin{equation}\label{sratioeqn}
\frac{\vert\tilde{B}_r\rvert^2}{2(\bm{k}\cdot\mvct{B})^2\tilde{U}}\frac{\partial^2 \tilde{U}}{\partial r^2}=\frac{w^4}{2^9 L^2 s^2}\approx \frac{w^2}{2^5 s^2}.
\end{equation}\par
From the ratio, we can see the first term here is dominate at small $|s|$ and can be ignored as $|s|$ becomes big. 
In the following, we will give the form of the parallel dynamo effect at different regions.

\subsubsection{\label{sec:rg1}  The region $|s|\ll 2^{-5/2}w$}
In this region, the ratio given by Eq.~(\ref{sratioeqn}) is large. 
Therefore, we can neglect the second term of Eq.~(\ref{ssffnlineqn})'s LHS, which gives
\begin{equation}
\fvct{B}\cdot\nabla\overline\lambda- \fvct{B}\cdot\nabla\left(\frac{\fvct{B}_\perp^*\cdot\nabla \tilde{U}}{2\eta\mAbsSqr{B}}\right)=0.
\end{equation}

As we mentioned before, the first term involves the dynamo driven parallel current density, $\delta\overline\lambda$. 
If we substitute $\delta\overline\lambda$ back into the above equation, we can find
\begin{equation}\label{eqn:fl}
\fvct{B}\cdot\nabla\left(\overline\lambda+\delta\overline\lambda\right)=0,
\end{equation}
which means the modified parallel current density, $\overline\lambda+\delta\overline\lambda$, is constant at this region. 

Eq.~(\ref{eqn:fl}) shows that no matter what the original $\overline\lambda$ profile is, the steady state parallel current density will be adjusted into constant by the dynamo driven $\delta\overline\lambda$.

\subsubsection{\label{sec:rg2}  The region  $|s|\gg 2^{-5/2}w$}
In this region, which is far from the rational surface, the first term of Eq.~(\ref{ssffnlineqn})'s LHS should be neglected. 
We can get
\begin{equation}
\fvct{B}\cdot\nabla\overline\lambda+\frac{(\bm k\cdot \mvct{B})^2 }{\eta \mAbsSqr{B}}\tilde U=0.
\end{equation}
Then, $\tilde{U}$ can be easily found as
\begin{equation}
\tilde U=-\frac{\eta \mAbsSqr{B}}{(\bm k\cdot \mvct{B})^2}\fvct{B}\cdot\nabla\overline\lambda.
\end{equation}
Recalling the expression of the parallel dynamo effect, Eq.~(\ref{dfeqn}), gives
\begin{equation}\label{eqn:lindf}
\varepsilon_\parallel=\frac{1}{2r\mAbs{B}}\frac{\partial}{\partial r}\left[\frac{\eta r\mAbsSqr{B}|{\tilde{B}_r}|^2}{(\bm k\cdot \mvct{B})^2 }\frac{\partial\overline\lambda}{\partial r}\right], 
\end{equation}
This is actually the same form as Strauss\cite{Strauss1985} found. 
It is not surprising because the first term's exclusion makes the treatment degenerate into the linear tearing mode theory.
And Strauss got the result using the linear tearing mode theory.

\subsubsection{The whole region}
The analytical solution of $\tilde{U}$ exists when the two terms are both considered simultaneously. 
Rewriting Eq.~(\ref{ssffnlineqn}) gives
\begin{equation}\label{seqn1}
\frac{\partial^2 \tilde{U}}{\partial r^2}-\frac{2(\bm k\cdot \mvct{B})^2}{\lvert\tilde{B}_r\rvert^2}\tilde{U}=\frac{2\eta\mAbsSqr{B}}{\lvert\tilde{B}_r\rvert}\frac{\partial \overline\lambda}{\partial r}.
\end{equation}
where we only keep the second order radial derivatives of $\tilde{U}$ in the first term of Eq.~(\ref{ssffnlineqn})'s LHS.

To illustrate the main effect and simplify the equation, we only take the first two terms of the $\overline\lambda$'s expansion, Eq.~(\ref{eqn:ld}),  into consideration. 
The form of $\bm k\cdot \mvct{B}$ is taken as 
$|\bm{k}||\mvct{B}|\overline\lambda_0 s$ as mentioned earlier. 
Then, Eq.~(\ref{seqn1}) is written in a more convenient form by introducing the characteristic length $d_1$ and new variables $X_1$ and $Y_1$ defined by
\begin{equation}\begin{split}\label{chalen}
&d_1=\left(\frac{|\tilde{B}_r|^2}{2|\bm{k}|^2|\mvct{B}|^2\overline\lambda_0^2}\right)^{1/4}=2^{-9/4}w,\\
&X_1=\frac{s}{d_1}\quad \text{and} \quad  Y_1=\left(\frac{2\eta\mAbsSqr{B}d_1^2}{\lvert\tilde{B}_r\rvert}\frac{\partial \overline\lambda}{\partial s}\Big|_{s=0}\right)^{-1}\tilde{U}.
\end{split}\end{equation}
Thus, substituting into Eq.~(\ref{seqn1}) gives
\begin{equation}\label{aseqn1}
\frac{\dif^2 Y_1}{\dif X_1^2}-X_1^2 Y_1=1
\end{equation}
According to Appendix~\ref{sec:appd}, the solution of the above second-order differential equation is
\begin{equation}\label{ysol1}
Y_1=-\frac{1}{2}\int_0^1 \dif \mu (1-\mu^2)^{-3/4}\exp{\left(-\frac{1}{2}\mu X_1^2\right)}.
\end{equation}
Actually, the solution is determined by the boundary condition. 
We assume the domain of $X_1$ is $(-\infty, \infty)$ and $Y_1=0$ at infinity for simplicity.

So that we can get the form of dynamo effect from Eqs.~(\ref{dfeqn}), (\ref{chalen}) and (\ref{ysol1}),
\begin{equation}
\varepsilon_\parallel =-\eta \mAbs{B} d_1 \frac{\partial \overline\lambda}{\partial s}\Big|_{s=0}\frac{\dif Y_1}{\dif X_1}.
\end{equation}
Obviously, $Y_1$ is an even function hence $\varepsilon_\parallel$ is an odd function. 
Therefore it drives opposite current density at different sides of the rational surface. 

The modification of $\overline\lambda$ induced by the parallel dynamo effect, 
\begin{equation}
\delta\overline\lambda=-\frac{\partial \overline\lambda}{\partial s}\Big|_{s=0}s\int_0^1 \dif \mu (1-\mu^2)^{-3/4}\frac{\mu}{2}\exp{\left(-\frac{\mu X_1^2 }{2}\right)}.
\end{equation}
The integration here approximately equal to 1 near $X_1=0$, which makes $\overline\lambda+\delta\overline\lambda\approx 0$ near the rational surface.

\begin{figure}[t]
    \includegraphics[scale=0.45]{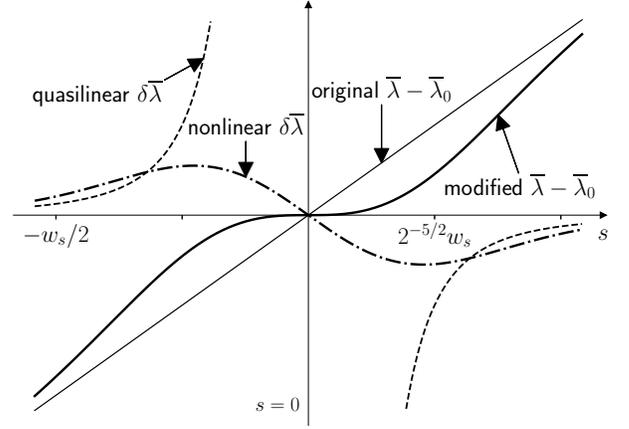}
    \caption{\label{fig:sssol}
    The $\delta\overline\lambda$ both in nonlinear and quasilinear form.The $\overline\lambda$ profile before and after modification by the nonlinear dynamo effect.}
\end{figure}

Fig.~\ref{fig:sssol} shows the nonlinear form of $\delta\overline\lambda$. 
The quasilinear $\delta\overline\lambda$ from Eq.~(\ref{eqn:lindf}) is also illustrated as a comparison. 
The original $\overline\lambda$ profile and modified  $\overline\lambda$ profile are also plotted.
The $\overline\lambda_0$ is deducted so the values equals 0 at $s=0$.
The ``saturated island width'', $w_\mathrm{s}$, and the separatrix, $|s|= 2^{-5/2}w_\mathrm{s}$, of the two regions Section (\ref{sec:rg1}) and (\ref{sec:rg2}) are also displayed.

The original $\overline\lambda$ profile is assumed as a linear function of $s$.
The modified $\overline\lambda$ profile is obtained by adding the nonlinear $\delta\overline\lambda$ into the original $\overline\lambda$.   
The original $\overline\lambda$ profile is almost entirely flattened by $\delta\overline\lambda$ near the rational surface. 
This is the current redistribution effect driven by the parallel dynamo effect. 
As for the external driven current concentrate at the edge, the dynamo effect may be useful in explaining the current transport from the edge to the core.

As the distance goes far from the rational surface, the nonlinear $\delta \overline\lambda$ becomes smaller and close to the quasilinear $\delta\overline\lambda$. 
The reason has been explained in Section (\ref{sec:rg2}).  

The above discussion shows the parallel dynamo effect's analytical expression under the simple linear distributed parallel current density profile.  
For an arbitrary parallel current density profile, the conclusions of Section (\ref{sec:rg1}) and (\ref{sec:rg2}) are still valid, but the solution of the whole region is more complicated.
A practical procedure is to perform Taylor expansion of $\overline\lambda$ and treat each term separately.
However, the first two terms of Taylor expansion are enough because of the narrowness of the region where the dynamo effect is essential.

\begin{figure}[tb]
    \includegraphics[scale=0.45]{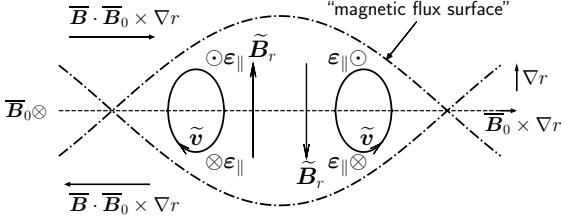}
    \caption{\label{fig:parallel_dynamo}The parallel dynamo effect generated by magnetic and velocity perturbation. }
\end{figure}

Fig.~\ref{fig:parallel_dynamo} demonstrates the direction of the parallel dynamo effect at different sides of the rational surface. 
The parallel dynamo effect, $\bm\varepsilon_\parallel$, is generated by $\fvct{B}_r$ and $\fvct{v}\cdot\mvct{B}_0\times\nabla r$, where $\mvct{B}_0$ is the mean magnetic field at the rational surface.

The mean current density is along the direction of $\mvct{B}_0$ at the whole region.
At the $r>r_s$ side, $\bm\varepsilon _\parallel$ is opposite to $\mvct{B}_0$ and reduces the mean current density.
At the $r<r_s$ side, $\bm\varepsilon_\parallel$ changes direction because of the reversal $\fvct{v}\cdot\mvct{B}_0\times\nabla r$, making the mean current density increase.
This clearly shows how the current redistribution occurs near the rational surface.

\subsection{Growth stage}
The time derivative terms are included in this section.
Some assumptions have to be made to simplify Eq.~(\ref{ffnlineqn}). 
Since $ \fvct{B}_\perp\approx \tilde{B}_r\nabla r$ and $\tilde{B}_r$ is constant, the second term of Eq.~(\ref{ffnlineqn})'s LHS becomes
\begin{equation}\label{fBpAeqn}
\fvct{B}\cdot\nabla\left(\frac{\fvct{B}_\perp^*}{2\eta\mAbsSqr{B}}\cdot\frac{\partial \fvct{A}}{\partial t}\right)=\frac{|\tilde{B}_r|^2}{2\mAbsSqr{B}}\frac{\mathrm{i}\bm k\cdot \mvct{B}}{\eta}\frac{\partial}{\partial t} \frac{\fvct{A}\cdot\mvct{B}}{\mAbsSqr{B}},
\end{equation}
where Eq.~(\ref{Areqn}) is used to replace $\tilde{A}_r$. 
Then, if the time derivative of $\mvct{A}$ is neglected, comparing the second and third terms of Eq.~(\ref{ffnlineqn})'s LHS, we can easily find the ratio is $|\tilde{B}_r|^2/2\mAbsSqr{B}$. 
Therefore, the second term can be ignored when compared to the third term of Eq.~(\ref{ffnlineqn})'s LHS.

Introducing the growth rate, $\gamma$, to represent $\partial /\partial t$, the simplified Eq.~(\ref{ffnlineqn}) can be written as
\begin{equation}\label{gsffnlineqn}
\frac{2\eta\rho\gamma+\Sqr{\tilde{B}_r}}{2\eta\mAbsSqr{B}}\frac{\partial^2\tilde{U}}{\partial s^2}-\frac{|\bm{k}|^2\overline\lambda_0^2}{\eta}s^2\tilde{U}=\tilde{B}_r\frac{\partial\overline\lambda}{\partial s}-\frac{\tilde{B}_r\overline\lambda_0\gamma}{\eta}s.
\end{equation}
This equation is the same as Rutherford's except that we keep the $\overline\lambda$ gradient term. 
Again, we only consider the first two terms of the $\overline\lambda$'s expansion. 
Then, introducing another characteristic length $d_2$ and new variables $X_2$ and $Y_2$ defined by  
\begin{equation}\begin{split}\label{chalen2}
&d_2=\left(\frac{2\eta\rho\gamma+|\tilde{B}_r|^2}{2|\bm{k}|^2\mAbsSqr{B}\overline\lambda_0^2}\right)^{1/4},\\%=\left(\frac{1}{|\bm{k}|^2 \overline\lambda_0^2}\frac{\tau_\mathrm{A}^2 \gamma}{\tau_\mathrm{R}}+\frac{w^4}{2^9}\right)^{1/4},\\
&X_2=\frac{s}{d_2},\quad \quad  Y_2=\left(\frac{2\eta\mAbsSqr{B}|\tilde{B}_r| d_2^2}{2\eta\rho\gamma+|\tilde{B}_r|^2}\frac{\partial \overline\lambda}{\partial s}\Big|_{s=0}\right)^{-1}\tilde{U},\\
&\text{and}\quad  C=\frac{\gamma d_2\overline\lambda_0}{\eta}\left(\frac{\partial \overline\lambda}{\partial s}\Big|_{s=0}\right)^{-1}.
\end{split}\end{equation}
Eq.~(\ref{gsffnlineqn}) can be written in a more convenient form,
\begin{equation}\label{aseqn2}
\frac{\dif^2 Y_2}{\dif X_2^2}-X_2^2 Y_2=1-CX_2.
\end{equation}

According to Appendix~\ref{sec:appd}, the solution is 
\begin{equation}\begin{split}\label{ysol2}
Y_2&=-\frac{1}{2}\int_0^1 \dif \mu (1-\mu^2)^{-3/4}\exp{\left(-\frac{1}{2}\mu X_2^2\right)}\\
&+\frac{CX_2}{2}\int_0^1 \dif \mu (1-\mu^2)^{-1/4}\exp{\left(-\frac{1}{2}\mu X_2^2\right)}.
\end{split}\end{equation}

The first term of Eq.~(\ref{ysol2})'s RHS is the same as Eq.~(\ref{ysol1}) and plays the role of flattening the current density. 
The second term of Eq.~(\ref{ysol2})'s RHS is the eddy current to slow down the growth of tearing mode as in nonlinear tearing mode theory\cite{Rutherford1973}.
The first term is an even part, and it does not affect the growth rate.
The second term actually changes the growth rate from the linear stage to the nonlinear stage.

If we include the higher order of $\overline\lambda$'s expansion, the second derivative of $\overline\lambda$ will contribute to the second term.
The growth rate may have a slight correction by this effect. 
% we will not discuss the detail in this paper.

From Eqs.~(\ref{dfeqn}), (\ref{chalen2}) and (\ref{ysol2}), the form of dynamo effect is
\begin{equation}\label{dfeqn2}
\varepsilon_\parallel =-\frac{\Sqr{\tilde{B}_r }}{2\eta\rho\gamma+\Sqr{\tilde{B}_r}}\eta\mAbs{B} d_2\frac{\partial \overline\lambda}{\partial s}\Big|_{s=0}\frac{\dif Y_2}{\dif X_2},
\end{equation}
Where the second term of Eq.~(\ref{dfeqn}) is omitted because of Eqs.~(\ref{fBpjeqn}) and (\ref{fBpAeqn}). 
The ratio of the two components in the denominator of Eq.~(\ref{dfeqn2}) is
\begin{equation}\label{eqn:ratio2}
\frac{2\eta\rho\gamma}{|\tilde{B}_r|^2}=\frac{2^9}{w^4 |\bm{k}|^2 \overline\lambda_0^2}\frac{\tau_\mathrm{A}^2 \gamma}{\tau_\mathrm{R}},
\end{equation}
where $\tau_\mathrm{A}=\rho^{1/2}a/|\mvct{B}|$ is the Alfv\'{e}n transit time, $\tau_\mathrm{R}=a^2/\eta$ is the resistive diffusion time, and $a$ is the minor radius. 
If this radio is small, the dynamo effect is essential.

At linear growth stage, the growth rate\cite{Wesson2011}
\begin{equation}
\gamma=0.55\left[(|\bm{k}|\overline\lambda_0)^{1/2}a^2\Delta'\right]^{4/5}\tau_\mathrm{R}^{-3/5}\tau_\mathrm{A}^{-2/5},
\end{equation}
where $\Delta'$ is an important parameter in tearing mode theory and is determined by outer region. 
The ratio becomes
\begin{equation}
\frac{2\eta\rho\gamma}{|\tilde{B}_r|^2}=282\left(\frac{a^2\Delta'}{|\bm{k}|^2\overline\lambda_0^2 w^5}\right)^{4/5} \left(\frac{\tau_\mathrm{R}}{\tau_\mathrm{A}}\right)^{-8/5}.
\end{equation}\par
As the ``island width'' is small at the beginning, the ratio is large.
However, the magnetic Reynolds number, $ \tau_\mathrm{R}/\tau_\mathrm{A}$, is prominent at most fusion plasmas.
For a typical $m=2$ mode, if we take $ \tau_\mathrm{R}/\tau_\mathrm{A}\approx10^6$, $a\Delta'\approx 10$, $|\bm{k}|\approx m/a$ and $\overline\lambda_0\approx 1/a$.
The condition for the ratio, Eq.~(\ref{eqn:ratio2}), being unity gives $w/a\approx 0.026$.
It is about a quarter of the normally saturated ``island width'', $w_\mathrm{s}/a=0.1$. 
The unity ratio means the dynamo effect is half of that in the steady state. 
We should expect the dynamo effect is essential even in the linear growth stage.

At the nonlinear growth stage, the growth rate becomes smaller as ``island width'' grows bigger. 
This makes the ratio, Eq.~(\ref{eqn:ratio2}), decrease much faster. 
The dynamo effect becomes significant at this stage.

\section{\label{sec:pressure} The pressure gradient}
The situation that includes the pressure gradient is considered in this section.
We give both the perpendicular and parallel dynamo effect of tearing modes. 
The perpendicular dynamo effect is first derived for the tearing modes.
The parallel dynamo effect has some difference from that in the force-free plasma.

Taking curl of the momentum equation, Eq.~(\ref{mmteqn}), to eliminate the $\nabla p$ and linearizing gives 
\begin{equation}\label{pnlineqn}
\overline\rho \frac{\partial}{\partial t} \nabla\times \fvct{v}=\mvct{B}\cdot\nabla \fvct{j}+\fvct{B}\cdot\nabla\mvct{j}-\mvct{j}\cdot\nabla \fvct{B}-\fvct{j}\cdot\nabla\mvct{B}.
\end{equation}\par
The elimination of $\nabla p$ avoids dealing with the $\nabla \tilde{p}$ term. 
The mean current density can be separated into two parts, parallel and perpendicular to the mean magnetic field respectively,
\begin{equation}\label{jeqn}
\mvct{j}=\overline{\lambda}\mvct{B}+\frac{\overline{j}_\perp}{\mAbs{B}}\mvct{B}\times\nabla r,
\end{equation}
The perpendicular current density, $\overline{j}_\perp$, can be represented in term of $\nabla \overline{p}$.
Using the mean part of force balance equation, Eq.~(\ref{mmteqn}), we get 
\begin{equation}\label{jpeqn}
\overline{j}_\perp=\mvct{j}\cdot(\mvct{B}\times\nabla r)=\frac{\partial \overline{p}}{\partial r}.
\end{equation}

\subsection{The perpendicular dynamo effect}
The $\nabla r$ component of Eq.~(\ref{pnlineqn}) is
\begin{equation}\label{rpeqn}
\overline\rho \frac{\partial}{\partial t} (\nabla r\cdot\nabla\times \fvct{v})=\mathrm{i}\bm{k}\cdot\mvct{B}\tilde{j}_r-\mathrm{i}\bm{k}\cdot\mvct{j}\tilde{B}_r.
\end{equation}
Substituting Eq.~(\ref{jeqn}) into above equation, and replacing $\overline j_\perp$ with $\overline{j}_\perp+\delta\overline{j}_\perp$.
Then Eq.~(\ref{rpeqn}) becomes
\begin{equation}\begin{split}\label{rpnlineqn}
&\overline\rho\frac{\partial}{\partial t}\left[\mathrm{i} \mvct{B}\cdot(\nabla r\times \bm{k})\frac{\fvct{v}\cdot\mvct{B}}{\mAbsSqr{B}}-\mathrm{i}\bm{k}\cdot\mvct{B}\frac{\fvct{v}\cdot(\mvct{B}\times\nabla r)}{\mAbsSqr{B}}\right] =\\
&\qquad\quad-\mathrm{i}\tilde{B}_r\bm{k}\cdot\mvct{B}(\overline\lambda-\overline\lambda_0)+\mathrm{i}\tilde{B}_r\mvct{B}\times\nabla r\cdot\bm{k}\frac{\overline{j}_\perp-\delta\overline{j}_\perp}{\mAbs{B}},
\end{split}\end{equation}
where $\tilde{j}_r$ is replaced using Eq.~(\ref{jreqn}).
From Eq.~(\ref{dfpeqn}), the dynamo induced perpendicular current density, $\delta\overline{j}_\perp$, is 
\begin{equation}\label{djpeqn}
\delta\overline{j}_\perp=\frac{\tilde{B}_r^*\fvct{v}\cdot\mvct{B}}{2\eta\mAbs{B}}-\frac{\fvct{B}^*\cdot\mvct{B}\tilde{\upsilon}_r}{2\eta\mAbs{B}}.
\end{equation}

If looking at the steady state, the form of $\delta\overline{j}_\perp$ is directly obtained from Eq.~(\ref{rpnlineqn}),
\begin{equation}
\delta\overline{j}_\perp=-\overline{j}_\perp-\overline\lambda_0\mAbs{B}\frac{\partial \overline\lambda}{\partial s}\Big|_{s=0}s^2,
\end{equation}
where $\overline\lambda$ has been Taylor expanded. 
The approximations, Eq.~(\ref{kBapprox}) and $\mvct{B}\times\nabla r\cdot\bm{k}\approx \Abs{k}\mAbs{B}$, are adopted here.

This form clearly shows that the current driven by the perpendicular dynamo effect offsets the mean perpendicular current density into zero near the rational surface. 
From Eq.~(\ref{jpeqn}), the pressure gradient is eliminated near the rational surface.
Again, the flux average is not used here.

If the growth state is considered, we need to know the relation of fluctuating velocity's three components to solve Eq.~(\ref{rpnlineqn}). 
For simplification, we neglect the second term of Eq.~(\ref{djpeqn})'s RHS, because $\tilde{B}_r$ is usually much greater than $\fvct{B}\cdot\mvct{B}/\mAbs{B}$. 
We also ignore the second term of Eq.~(\ref{rpnlineqn} )'s LHS, as $\bm{k}\cdot\mvct{B}\ll \mvct{B}\cdot(\nabla r\times \bm{k})$. 
Then, Eq.~(\ref{rpnlineqn}) only involves $\fvct{v}\cdot\mvct{B}$, and becomes
\begin{equation}
\frac{2\overline\rho\gamma\eta+\Sqr{\tilde{B}_r}}{2\eta\mAbsSqr{B}}\fvct{v}\cdot\mvct{B}=-\tilde{B}_r\overline\lambda_0\frac{\partial \overline\lambda}{\partial s}\Big|_{s=0}s^2-\tilde{B}_r\frac{\overline{j}_\perp}{\mAbs{B}}.
\end{equation}
So that $\fvct{v}\cdot\mvct{B}$ can be easily calculated as
\begin{equation}
\fvct{v}\cdot\mvct{B}=-\frac{2\eta\mAbsSqr{B}\tilde{B}_r}{2\overline\rho\gamma\eta+\Sqr{\tilde{B}_r}}\left(\frac{\overline{j}_\perp}{\mAbs{B}}+\overline\lambda_0\frac{\partial \overline\lambda}{\partial s}\Big|_{s=0}s^2\right),
\end{equation}
The dynamo induced perpendicular current density is
\begin{equation}
\delta \overline{j}_\perp=-\frac{\Sqr{\tilde{B}_r}}{2\overline\rho\gamma\eta+\Sqr{\tilde{B}_r}}\left(\overline{j}_\perp+\overline\lambda_0\mAbs{B}\frac{\partial \overline\lambda}{\partial s}\Big|_{s=0}s^2\right).
\end{equation}\par
When tearing mode grows, the magnetic perturbation becomes bigger while the growth rate becomes smaller.
$\delta\overline{j}_\perp$ becomes more approaches the steady state and the pressure more flatten. 
The coefficient here is the same as that in Eq.~(\ref{dfeqn2}). 

\begin{figure}
    \includegraphics[scale=0.45]{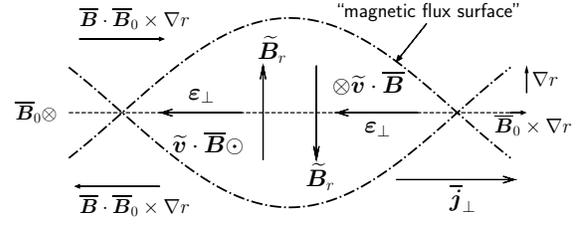}
    \caption{\label{fig:perpendicular_dynamo}The perpendicular dynamo effect generated by magnetic and velocity perturbation. }
\end{figure}
Fig.~\ref{fig:perpendicular_dynamo} demonstrates that the direction of the perpendicular dynamo is opposite to $\mvct{j}_\perp$. 
The generation of $\bm\varepsilon_\perp$ requires $\fvct{B}_r$ and $\fvct{v}\cdot\mvct{B}$. 
It is different with the force-free situation where $\fvct{v}\cdot\mvct{B}$ is not needed.
This means the velocity perturbation will develop a 3D structure when the pressure gradient is included.

\subsection{The parallel dynamo effect}
Taking $\mvct{B}$ component of Eq.~(\ref{pnlineqn}) leads to
\begin{equation}\begin{split}\label{Bpnlineqn}
\overline\rho \frac{\partial}{\partial t} (\mvct{B}\cdot\nabla\times \fvct{v})&=\mathrm{i}\bm{k}\cdot\mvct{B}(\fvct{j}\cdot\mvct{B})-\mathrm{i}\bm{k}\cdot\mvct{j}(\fvct{B}\cdot\mvct{B})\\
&+\tilde{B}_r\left(\mvct{B}\cdot\frac{\partial \mvct{j}}{\partial r}\right)-\tilde{j}_r\left(\mvct{B}\cdot\frac{\partial \mvct{B}}{\partial r}\right).
\end{split}\end{equation}

The divergence of linearized Eq.~(\ref{mohmeqn}) gives
\begin{equation}\label{d2Ueqn}
\mvct{B}\cdot\nabla\times\fvct{v}=\nabla^2\tilde{U}-\fvct{v}\cdot\mvct{j}\approx \frac{\partial^2\tilde{U}}{\partial r^2},
\end{equation}
where we only keep the dominant second radial derivative term.
Because  $\fvct{v}\sim\nabla \tilde{U}\times\mvct{B}/\mAbs{B}$ and the resistive layer is narrow.

The $\mvct{B}$ component of linearized Eq.~(\ref{mohmeqn}) gives
\begin{equation}\label{j1Beqn}
\fvct{j}\cdot\mvct{B}=-\frac{1}{\eta}\left(\mathrm{i}\bm{k}\cdot\mvct{B}\tilde{U}+\frac{\partial \fvct{A}\cdot \mvct{B}}{\partial t}\right),
\end{equation}
where the time derivative of $\mvct{B}$ is ignored.
Substituting Eqs.~(\ref{d2Ueqn}) and (\ref{j1Beqn}) to Eq.~(\ref{Bpnlineqn}).
Utilizing the approximations, Eqs.~(\ref{kBapprox}), (\ref{jreqn}) and $\mvct{B}\times\nabla r\cdot\bm{k}\approx \Abs{k}\mAbs{B}$, we get the equation for $\tilde{U}$ after some algebra,
\begin{equation}\begin{split}
&\overline\rho\gamma\frac{\partial^2 \tilde{U}}{\partial s^2}-\frac{\AbsSqr{k}\mAbsSqr{B}\overline\lambda_0^2}{\eta}s^2\tilde{U}=\\
&\qquad\qquad\tilde{B}_r\mAbsSqr{B}\left(\frac{\partial\overline\lambda}{\partial r}-2\frac{\overline\lambda_0}{\mAbsSqr{B}}\frac{\partial \overline p}{\partial r}\right)-\frac{\gamma\tilde{B}_r\mAbsSqr{B}\overline\lambda_0}{\eta}s,
\end{split}\end{equation}
where the time derivative is represented as $\gamma$. 
Again, to include the nonlinear term, we replace $\overline\lambda$ with $\overline\lambda+\delta\overline\lambda$. 
Finally we can get
\begin{equation}\begin{split}\label{pnlineqn2}
&\frac{2\eta\rho\gamma+\Sqr{\tilde{B}_r}}{2\eta\mAbsSqr{B}}\frac{\partial^2\tilde{U}}{\partial s^2}-\frac{|\bm{k}|^2\overline\lambda_0^2}{\eta}s^2\tilde{U}=\\
&\qquad\qquad\qquad\quad\tilde{B}_r\left(\frac{\partial\overline\lambda}{\partial s}-2\frac{\overline\lambda_0}{\mAbsSqr{B}}\frac{\partial \overline p}{\partial s}\right)\Bigg|_{s=0}-\frac{\tilde{B}_r\overline\lambda_0\gamma}{\eta}s,
\end{split}\end{equation}
where $\overline\lambda$ and $\overline p$ are Taylor expanded, and we only keep the leading terms.  
Except for the first term, this form is the same as Eq.~(\ref{gsffnlineqn}). 
If the pressure gradient is neglected as in force-free plasma, this equation will degenerate into Eq.~(\ref{gsffnlineqn}). 
Therefore, the solution can be obtained conveniently by the replacement,
\begin{equation}
\frac{\partial \overline\lambda}{\partial s}\Big|_{s=0}\rightarrow\left(\frac{\partial\overline\lambda}{\partial s}-2\frac{\overline\lambda_0}{\mAbsSqr{B}}\frac{\partial \overline p}{\partial s}\right)\Bigg|_{s=0},
\end{equation}
in $Y_2$ and $C$. We introduce $Y_2^\prime$ and $C'$,
\begin{equation}\begin{split}
&Y_2^\prime=\left[\frac{2\eta\mAbsSqr{B}|\tilde{B}_r| d_2^2}{2\eta\rho\gamma+\Sqr{\tilde{B}_r}}\left(\frac{\partial \overline\lambda}{\partial s}-\frac{2\overline\lambda_0}{\mAbsSqr{B}}\frac{\partial \overline p}{\partial s}\right)\Bigg|_{s=0}\right]^{-1}\tilde{U},\\
&C'=\frac{\gamma d_2\overline\lambda_0}{\eta}\left[\left(\frac{\partial \overline\lambda}{\partial s}-\frac{2\overline\lambda_0}{\mAbsSqr{B}}\frac{\partial \overline p}{\partial s}\right)\Bigg|_{s=0}\right]^{-1}.
\end{split}\end{equation}
So that, the solution of $Y_2^\prime$ is the same as $Y_2$ when replacing $C$ by $C'$ in Eq.~(\ref{ysol2}).\par
The parallel dynamo effect is calculated as
\begin{equation}\label{pdfeqn2}
\varepsilon_\parallel =-\frac{ \eta\mAbs{B}\Sqr{\tilde{B}_r} d_2}{2\eta\rho\gamma+\Sqr{\tilde{B}_r}}\left(\frac{\partial \overline\lambda}{\partial s}-\frac{2\overline\lambda_0}{\mAbsSqr{B}}\frac{\partial \overline p}{\partial s}\right)\Bigg|_{s=0}\frac{\dif Y_2^\prime}{\dif X_2}.
\end{equation}
If the relation, $p+\mAbsSqr{B}/2=\text{constant}$, is used, 
\begin{equation}
\frac{\partial\overline\lambda}{\partial s}-2\frac{\overline\lambda_0}{\mAbsSqr{B}}\frac{\partial \overline p}{\partial s}\approx \frac{1}{\mAbsSqr{B}}\frac{\partial\mvct{j}\cdot\mvct{B}}{\partial s}.
\end{equation}
This seems that the parallel dynamo effect will flatten $\mvct{j}\cdot\mvct{B}$ profile instead of $\overline\lambda$ profile. 
This is contradicting with Taylor's theory, where the tendency of flattening $\overline\lambda$ is required. 
Taylor's theory is suitable in the force-free plasma, while our result is derived with the pressure gradient included. 
This may give an idea of extending Taylor's theory into the plasma with a pressure gradient. 
% On the other hand, if combining the parallel and perpendicular dynamo effect, we will get a force-free area near the rational surface.
% will eliminate the pressure gradient as we discussed in the previous section.
%which makes $\mAbs{B}=\text{constant}$. 
%This means the plasma becomes force-free near the rational surface, and the flattening effect of $\overline\lambda$ is still valid.

\section{\label{sec:conclusion} conclusions and discussion}

In this paper, the nonlinear dynamo effect of the tearing modes is derived in the frame of the resistive MHD model. 
The dynamo effect parallel to $\mvct{B}$ was considered both in the force-free plasma and the plasma with pressure gradient included.
The dynamo effect perpendicular to $\mvct{B}$ was first derived in the plasma with pressure gradient. 

In the force-free plasma, it was found that the parallel dynamo effect would drive opposite current densities at different sides of the rational surface. 
Thus the $\overline\lambda$ profile is flattened by the dynamo effect near the rational surface.
The fast transport along the magnetic flux surface is not used here.
For the turbulent plasma, there are many rational surfaces, so the $\overline\lambda$ profile is flattened in the entire turbulent region.
This provided a reasonable explanation of the Taylor relaxation process through the dynamo effect.

The quasilinear theory also indicated that the parallel dynamo effect produced the flattening effect of the current density. 
However, it was only suitable for the region that far from the rational surface, and the flattening effect was minimal.
This paper's nonlinear theory gave the proper form of the parallel dynamo effect in the entire region.
It was found that the flattening effect mainly influences the region near the rational surface. 
At the region far from the rational surface, the parallel dynamo effect is relatively small. 
Moreover, our nonlinear form of the parallel dynamo effect is the same as the quasilinear form under the approximation that the distance is far from the rational surface.
The reason is the nonlinear term can be ignored at the region far from the rational surface.

In the plasma with the pressure gradient, it was found that the perpendicular dynamo effect would drive a current density opposite to the mean perpendicular current density.
This eliminates the mean perpendicular current density, which indicates the elimination of the pressure gradient.

Besides, with the pressure gradient included, the parallel dynamo effect is found different from that in force-free plasma.
The parallel dynamo effect flattens $\mvct{j}\cdot\mvct{B}$ instead of $\overline\lambda$ when considering the pressure gradient. 
This may give inspiration for extending the Taylor relaxation theory with the pressure gradient included.

Most tearing mode theories had assumed the flatten current density using the flux average in the magnetic island.
Some tearing mode theories naturally ignored the current density gradient in the inner region like Li\cite{Li1995} did.
Our result gave another explanation of how the current density flattened in the magnetic island without using flux average. 
This indicates that the current density flattening assumption has a solid foundation in the tearing mode theory.
In addition, the tearing mode theories considering the current density gradient perpendicular to the magnetic flux may take this dynamo-induced flattening effect into consideration.

The turbulent plasma was decomposed into many tearing modes in this paper. 
Although we treated the tearing modes separately, the interactions of the tearing modes, especially the adjacent ones, should be considered in future work.

\begin{acknowledgments}
We wish to acknowledge Dr. Zhuping's team's support with the using of NIMROD and offering suggestions. We also thanks Dr. Huang Wenlong for the discussion about tearing mode. This work was supported by National Natural Science Foundation of China (Nos. 11827810 and 11875177), International Atomic Energy Agency Research Contract No. 22733.
\end{acknowledgments}

\appendix

\section{\label{sec:appd}The solution of $\frac{\dif^2 y}{\dif x^2}-x^2y=a+bx$}
Note that $a$ and $b$ are arbitrary real numbers here, do not confuse with minor radius mentioned above. This is a linear equation and can be separated into two equations:
\begin{align}
\frac{\dif^2 y}{\dif x^2}-x^2y=1,\label{eqn:A1}\\
\frac{\dif^2 y}{\dif x^2}-x^2y=x.\label{eqn:A2}
\end{align}
The solutions of above two equations are even and odd functions respectively. Therefore, we can deal with $x\geq 0$ first and get another part using symmetricity. Taking substitutions with
\begin{equation}\label{eqn:subs}
z=\frac{1}{2x^2},\quad u=z^{-1/4}y.
\end{equation}
Then, the equations becomes
\begin{align}
&z^2\frac{\dif^2 u}{\dif z^2}+z\frac{\dif u}{\dif z}-\left(z^2+\frac{1}{16}\right)u=\frac{1}{2}z^{3/4},\\
&z^2\frac{\dif^2 u}{\dif z^2}+z\frac{\dif u}{\dif z}-\left(z^2+\frac{1}{16}\right)u=\frac{\sqrt 2}{2}z^{5/4}.
\end{align}
These are two special cases of the non-homogeneous Bessel's differential equation, and the solutions are the second-kind version of modified Struve functions\cite{abramowitz1972handbook}. The two modified Struve functions here can be represented in term of the Poisson’s integral:
\begin{align}
&u(z)=-\frac{1}{2}z^{-1/4}\int_0^1(1-t^2)^{-3/4}\exp{(-zt)}\dif t,\\
&u(z)=-\frac{\sqrt{2}}{2}z^{1/4}\int_0^1(1-t^2)^{-1/4}\exp{(-zt)}\dif t.
\end{align}
Substituting back with Eq.~(\ref{eqn:subs}) and using the symmetricity, the solutions of Eqs.~(\ref{eqn:A1}) and (\ref{eqn:A2}) are given as
\begin{align}
&y(x)=-\frac{1}{2}\int_0^1 \dif \mu (1-\mu^2)^{-3/4}\exp{\left(-\frac{1}{2}\mu x^2\right)},\\
&y(x)=-\frac{x}{2}\int_0^1 \dif \mu (1-\mu^2)^{-1/4}\exp{\left(-\frac{1}{2}\mu x^2\right)}.
\end{align}
Combine the above two solutions, we can get the solution of the original equation:
\begin{equation}\begin{split}\label{eqn:sol}
y(x)&=-\frac{a}{2}\int_0^1 \dif \mu (1-\mu^2)^{-3/4}\exp{\left(-\frac{1}{2}\mu x^2\right)}\\
&-\frac{bx}{2}\int_0^1 \dif \mu (1-\mu^2)^{-1/4}\exp{\left(-\frac{1}{2}\mu x^2\right)}.
\end{split}\end{equation}

% If in two-column mode, this environment will change to single-column format so that long equations can be displayed. 
%\begin{widetext}
%$$\mbox{put long equation here}$$
%\end{widetext}

% Figures should be put into the text as floats. 
% Use the graphics or graphicx packages (distributed with LaTeX2e).
% Put the label that you will use with \ref{} command in the braces of the \label{} command.
%
% \begin{figure}
% \end{figure}

% Tables may be put in the text as floats.
% \begin{table}
% \end{tabular}
% \end{table}

%\begin{acknowledgments}
% Put your acknowledgments here.
%\end{acknowledgments}

\bibliography{dynamo-effect}

\providecommand{\noopsort}[1]{}\providecommand{\singleletter}[1]{#1}%
\begin{thebibliography}{25}%
\makeatletter
\providecommand \@ifxundefined [1]{%
 \@ifx{#1\undefined}
}%
\providecommand \@ifnum [1]{%
 \ifnum #1\expandafter \@firstoftwo
 \else \expandafter \@secondoftwo
 \fi
}%
\providecommand \@ifx [1]{%
 \ifx #1\expandafter \@firstoftwo
 \else \expandafter \@secondoftwo
 \fi
}%
\providecommand \natexlab [1]{#1}%
\providecommand \enquote  [1]{``#1''}%
\providecommand \bibnamefont  [1]{#1}%
\providecommand \bibfnamefont [1]{#1}%
\providecommand \citenamefont [1]{#1}%
\providecommand \href@noop [0]{\@secondoftwo}%
\providecommand \href [0]{\begingroup \@sanitize@url \@href}%
\providecommand \@href[1]{\@@startlink{#1}\@@href}%
\providecommand \@@href[1]{\endgroup#1\@@endlink}%
\providecommand \@sanitize@url [0]{\catcode `\\12\catcode `\$12\catcode
  `\&12\catcode `\#12\catcode `\^12\catcode `\_12\catcode `\%12\relax}%
\providecommand \@@startlink[1]{}%
\providecommand \@@endlink[0]{}%
\providecommand \url  [0]{\begingroup\@sanitize@url \@url }%
\providecommand \@url [1]{\endgroup\@href {#1}{\urlprefix }}%
\providecommand \urlprefix  [0]{URL }%
\providecommand \Eprint [0]{\href }%
\providecommand \doibase [0]{http://dx.doi.org/}%
\providecommand \selectlanguage [0]{\@gobble}%
\providecommand \bibinfo  [0]{\@secondoftwo}%
\providecommand \bibfield  [0]{\@secondoftwo}%
\providecommand \translation [1]{[#1]}%
\providecommand \BibitemOpen [0]{}%
\providecommand \bibitemStop [0]{}%
\providecommand \bibitemNoStop [0]{.\EOS\space}%
\providecommand \EOS [0]{\spacefactor3000\relax}%
\providecommand \BibitemShut  [1]{\csname bibitem#1\endcsname}%
\let\auto@bib@innerbib\@empty
%</preamble>
\bibitem [{\citenamefont {Moffatt}(1978)}]{Moffatt1984}%
  \BibitemOpen
  \bibfield  {author} {\bibinfo {author} {\bibfnamefont {H.~K.}\ \bibnamefont
  {Moffatt}},\ }\href {\doibase 10.1088/0031-9112/35/10/037} {\emph {\bibinfo
  {title} {Cambridge University Press}}}\ (\bibinfo {year} {1978})\BibitemShut
  {NoStop}%
\bibitem [{\citenamefont {Sokoloff}, \citenamefont {Stepanov},\ and\
  \citenamefont {Frick}(2014)}]{Sokoloff2014}%
  \BibitemOpen
  \bibfield  {author} {\bibinfo {author} {\bibfnamefont {D.~D.}\ \bibnamefont
  {Sokoloff}}, \bibinfo {author} {\bibfnamefont {R.~A.}\ \bibnamefont
  {Stepanov}}, \ and\ \bibinfo {author} {\bibfnamefont {P.~G.}\ \bibnamefont
  {Frick}},\ }\bibfield  {title} {\enquote {\bibinfo {title} {{Dynamos: from an
  astrophysical model to laboratory experiments}},}\ }\href {\doibase
  10.3367/ufne.0184.201403g.0313} {\bibfield  {journal} {\bibinfo  {journal}
  {Physics-Uspekhi}\ }\textbf {\bibinfo {volume} {57}},\ \bibinfo {pages}
  {292--311} (\bibinfo {year} {2014})}\BibitemShut {NoStop}%
\bibitem [{\citenamefont {Yoshizawa}(1990)}]{Yoshizawa1990}%
  \BibitemOpen
  \bibfield  {author} {\bibinfo {author} {\bibfnamefont {A.}~\bibnamefont
  {Yoshizawa}},\ }\bibfield  {title} {\enquote {\bibinfo {title}
  {{Self-consistent turbulent dynamo modeling of reversed field pinches and
  planetary magnetic fields}},}\ }\href {\doibase 10.1063/1.859484} {\bibfield
  {journal} {\bibinfo  {journal} {Physics of Fluids B}\ }\textbf {\bibinfo
  {volume} {2}},\ \bibinfo {pages} {1589--1600} (\bibinfo {year}
  {1990})}\BibitemShut {NoStop}%
\bibitem [{\citenamefont {Brandenburg}(2001)}]{Brandenburg2001}%
  \BibitemOpen
  \bibfield  {author} {\bibinfo {author} {\bibfnamefont {A.}~\bibnamefont
  {Brandenburg}},\ }\bibfield  {title} {\enquote {\bibinfo {title} {{The
  Inverse Cascade and Nonlinear Alpha‐Effect in Simulations of Isotropic
  Helical Hydromagnetic Turbulence}},}\ }\href {\doibase 10.1086/319783}
  {\bibfield  {journal} {\bibinfo  {journal} {The Astrophysical Journal}\
  }\textbf {\bibinfo {volume} {550}},\ \bibinfo {pages} {824--840} (\bibinfo
  {year} {2001})},\ \Eprint {http://arxiv.org/abs/0006186} {arXiv:0006186
  [astro-ph]} \BibitemShut {NoStop}%
\bibitem [{\citenamefont {Hokin}\ \emph {et~al.}(1991)\citenamefont {Hokin},
  \citenamefont {Almagri}, \citenamefont {Assadi}, \citenamefont {Beckstead},
  \citenamefont {Chartas}, \citenamefont {Crocker}, \citenamefont {Cudzinovic},
  \citenamefont {{Den Hartog}}, \citenamefont {Dexter}, \citenamefont {Holly},
  \citenamefont {Prager}, \citenamefont {Rempel}, \citenamefont {Sarff},
  \citenamefont {Scime}, \citenamefont {Shen}, \citenamefont {Spragins},
  \citenamefont {Sprott}, \citenamefont {Starr}, \citenamefont {Stoneking},
  \citenamefont {Watts},\ and\ \citenamefont {Nebel}}]{Hokin1991}%
  \BibitemOpen
  \bibfield  {author} {\bibinfo {author} {\bibfnamefont {S.}~\bibnamefont
  {Hokin}}, \bibinfo {author} {\bibfnamefont {A.}~\bibnamefont {Almagri}},
  \bibinfo {author} {\bibfnamefont {S.}~\bibnamefont {Assadi}}, \bibinfo
  {author} {\bibfnamefont {J.}~\bibnamefont {Beckstead}}, \bibinfo {author}
  {\bibfnamefont {G.}~\bibnamefont {Chartas}}, \bibinfo {author} {\bibfnamefont
  {N.}~\bibnamefont {Crocker}}, \bibinfo {author} {\bibfnamefont
  {M.}~\bibnamefont {Cudzinovic}}, \bibinfo {author} {\bibfnamefont
  {D.}~\bibnamefont {{Den Hartog}}}, \bibinfo {author} {\bibfnamefont
  {R.}~\bibnamefont {Dexter}}, \bibinfo {author} {\bibfnamefont
  {D.}~\bibnamefont {Holly}}, \bibinfo {author} {\bibfnamefont
  {S.}~\bibnamefont {Prager}}, \bibinfo {author} {\bibfnamefont
  {T.}~\bibnamefont {Rempel}}, \bibinfo {author} {\bibfnamefont
  {J.}~\bibnamefont {Sarff}}, \bibinfo {author} {\bibfnamefont
  {E.}~\bibnamefont {Scime}}, \bibinfo {author} {\bibfnamefont
  {W.}~\bibnamefont {Shen}}, \bibinfo {author} {\bibfnamefont {C.}~\bibnamefont
  {Spragins}}, \bibinfo {author} {\bibfnamefont {C.}~\bibnamefont {Sprott}},
  \bibinfo {author} {\bibfnamefont {G.}~\bibnamefont {Starr}}, \bibinfo
  {author} {\bibfnamefont {M.}~\bibnamefont {Stoneking}}, \bibinfo {author}
  {\bibfnamefont {C.}~\bibnamefont {Watts}}, \ and\ \bibinfo {author}
  {\bibfnamefont {R.}~\bibnamefont {Nebel}},\ }\bibfield  {title} {\enquote
  {\bibinfo {title} {{Global confinement and discrete dynamo activity in the
  MST reversed-field pinch}},}\ }\href {\doibase 10.1063/1.859642} {\bibfield
  {journal} {\bibinfo  {journal} {Physics of Fluids B}\ }\textbf {\bibinfo
  {volume} {3}},\ \bibinfo {pages} {2241--2246} (\bibinfo {year}
  {1991})}\BibitemShut {NoStop}%
\bibitem [{\citenamefont {Al-Karkhy}\ \emph {et~al.}(1993)\citenamefont
  {Al-Karkhy}, \citenamefont {Browning}, \citenamefont {Cunningham},
  \citenamefont {Gee},\ and\ \citenamefont {Rusbridge}}]{Al-Karkhy1993}%
  \BibitemOpen
  \bibfield  {author} {\bibinfo {author} {\bibfnamefont {A.}~\bibnamefont
  {Al-Karkhy}}, \bibinfo {author} {\bibfnamefont {P.~K.}\ \bibnamefont
  {Browning}}, \bibinfo {author} {\bibfnamefont {G.}~\bibnamefont
  {Cunningham}}, \bibinfo {author} {\bibfnamefont {S.~J.}\ \bibnamefont {Gee}},
  \ and\ \bibinfo {author} {\bibfnamefont {M.~G.}\ \bibnamefont {Rusbridge}},\
  }\bibfield  {title} {\enquote {\bibinfo {title} {{Observations of the
  magnetohydrodynamic dynamo effect in a spheromak plasma}},}\ }\href {\doibase
  10.1103/PhysRevLett.70.1814} {\bibfield  {journal} {\bibinfo  {journal}
  {Physical Review Letters}\ }\textbf {\bibinfo {volume} {70}},\ \bibinfo
  {pages} {1814--1817} (\bibinfo {year} {1993})}\BibitemShut {NoStop}%
\bibitem [{\citenamefont {Jarboe}\ \emph {et~al.}(2012)\citenamefont {Jarboe},
  \citenamefont {Victor}, \citenamefont {Nelson}, \citenamefont {Hansen},
  \citenamefont {Akcay}, \citenamefont {Ennis}, \citenamefont {Hicks},
  \citenamefont {Hossack}, \citenamefont {Marklin},\ and\ \citenamefont
  {Smith}}]{Jarboe2012}%
  \BibitemOpen
  \bibfield  {author} {\bibinfo {author} {\bibfnamefont {T.~R.}\ \bibnamefont
  {Jarboe}}, \bibinfo {author} {\bibfnamefont {B.~S.}\ \bibnamefont {Victor}},
  \bibinfo {author} {\bibfnamefont {B.~A.}\ \bibnamefont {Nelson}}, \bibinfo
  {author} {\bibfnamefont {C.~J.}\ \bibnamefont {Hansen}}, \bibinfo {author}
  {\bibfnamefont {C.}~\bibnamefont {Akcay}}, \bibinfo {author} {\bibfnamefont
  {D.~A.}\ \bibnamefont {Ennis}}, \bibinfo {author} {\bibfnamefont {N.~K.}\
  \bibnamefont {Hicks}}, \bibinfo {author} {\bibfnamefont {A.~C.}\ \bibnamefont
  {Hossack}}, \bibinfo {author} {\bibfnamefont {G.~J.}\ \bibnamefont
  {Marklin}}, \ and\ \bibinfo {author} {\bibfnamefont {R.~J.}\ \bibnamefont
  {Smith}},\ }\bibfield  {title} {\enquote {\bibinfo {title} {{Imposed-dynamo
  current drive}},}\ }\href {\doibase 083017} {\bibfield  {journal} {\bibinfo
  {journal} {Nuclear Fusion}\ }\textbf {\bibinfo {volume} {52}} (\bibinfo
  {year} {2012}),\ 083017}\BibitemShut {NoStop}%
\bibitem [{\citenamefont {Taylor}(1986)}]{Taylor1986}%
  \BibitemOpen
  \bibfield  {author} {\bibinfo {author} {\bibfnamefont {J.~B.}\ \bibnamefont
  {Taylor}},\ }\bibfield  {title} {\enquote {\bibinfo {title} {{Relaxation and
  magnetic reconnection in plasmas}},}\ }\href {\doibase
  10.1103/RevModPhys.58.741} {\bibfield  {journal} {\bibinfo  {journal}
  {Reviews of Modern Physics}\ }\textbf {\bibinfo {volume} {58}},\ \bibinfo
  {pages} {741--763} (\bibinfo {year} {1986})}\BibitemShut {NoStop}%
\bibitem [{\citenamefont {Strauss}(1985)}]{Strauss1985}%
  \BibitemOpen
  \bibfield  {author} {\bibinfo {author} {\bibfnamefont {H.~R.}\ \bibnamefont
  {Strauss}},\ }\bibfield  {title} {\enquote {\bibinfo {title} {{The dynamo
  effect in fusion plasmas}},}\ }\href {\doibase 10.1063/1.865238} {\bibfield
  {journal} {\bibinfo  {journal} {Physics of Fluids}\ }\textbf {\bibinfo
  {volume} {28}},\ \bibinfo {pages} {2786--2792} (\bibinfo {year}
  {1985})}\BibitemShut {NoStop}%
\bibitem [{\citenamefont {Bhattacharjee}\ and\ \citenamefont
  {Hameiri}(1986)}]{Bhattacharjee1986}%
  \BibitemOpen
  \bibfield  {author} {\bibinfo {author} {\bibfnamefont {A.}~\bibnamefont
  {Bhattacharjee}}\ and\ \bibinfo {author} {\bibfnamefont {E.}~\bibnamefont
  {Hameiri}},\ }\bibfield  {title} {\enquote {\bibinfo {title}
  {{Self-consistent dynamolike activity in turbulent plasmas}},}\ }\href
  {\doibase 10.1103/PhysRevLett.57.206} {\bibfield  {journal} {\bibinfo
  {journal} {Physical Review Letters}\ }\textbf {\bibinfo {volume} {57}},\
  \bibinfo {pages} {206--209} (\bibinfo {year} {1986})}\BibitemShut {NoStop}%
\bibitem [{\citenamefont {Ji}\ and\ \citenamefont {Prager}(2002)}]{Ji2002}%
  \BibitemOpen
  \bibfield  {author} {\bibinfo {author} {\bibfnamefont {H.}~\bibnamefont
  {Ji}}\ and\ \bibinfo {author} {\bibfnamefont {S.~C.}\ \bibnamefont
  {Prager}},\ }\bibfield  {title} {\enquote {\bibinfo {title} {{The $\alpha$
  dynamo effects in laboratory plasmas}},}\ }\href {\doibase
  10.22364/mhd.38.1-2.15} {\bibfield  {journal} {\bibinfo  {journal}
  {Magnetohydrodynamics}\ }\textbf {\bibinfo {volume} {38}},\ \bibinfo {pages}
  {191--210} (\bibinfo {year} {2002})}\BibitemShut {NoStop}%
\bibitem [{\citenamefont {Cappello}, \citenamefont {Bonfiglio},\ and\
  \citenamefont {Escande}(2006)}]{Cappello2006}%
  \BibitemOpen
  \bibfield  {author} {\bibinfo {author} {\bibfnamefont {S.}~\bibnamefont
  {Cappello}}, \bibinfo {author} {\bibfnamefont {D.}~\bibnamefont {Bonfiglio}},
  \ and\ \bibinfo {author} {\bibfnamefont {D.~F.}\ \bibnamefont {Escande}},\
  }\bibfield  {title} {\enquote {\bibinfo {title} {{Magnetohydrodynamic dynamo
  in reversed field pinch plasmas: Electrostatic drift nature of the dynamo
  velocity field}},}\ }\href {\doibase 10.1063/1.2177198} {\bibfield  {journal}
  {\bibinfo  {journal} {Physics of Plasmas}\ }\textbf {\bibinfo {volume}
  {13}},\ \bibinfo {pages} {56102} (\bibinfo {year} {2006})}\BibitemShut
  {NoStop}%
\bibitem [{\citenamefont {Mirnov}, \citenamefont {Hegna},\ and\ \citenamefont
  {Prager}(2004)}]{Mirnov2004}%
  \BibitemOpen
  \bibfield  {author} {\bibinfo {author} {\bibfnamefont {V.~V.}\ \bibnamefont
  {Mirnov}}, \bibinfo {author} {\bibfnamefont {C.~C.}\ \bibnamefont {Hegna}}, \
  and\ \bibinfo {author} {\bibfnamefont {S.~C.}\ \bibnamefont {Prager}},\
  }\bibfield  {title} {\enquote {\bibinfo {title} {{Two-fluid tearing
  instability in force-free magnetic configuration}},}\ }\href {\doibase
  10.1063/1.1773778} {\bibfield  {journal} {\bibinfo  {journal} {Physics of
  Plasmas}\ }\textbf {\bibinfo {volume} {11}},\ \bibinfo {pages} {4468}
  (\bibinfo {year} {2004})}\BibitemShut {NoStop}%
\bibitem [{\citenamefont {Bonfiglio}, \citenamefont {Cappello},\ and\
  \citenamefont {Escande}(2005)}]{Bonfiglio2005}%
  \BibitemOpen
  \bibfield  {author} {\bibinfo {author} {\bibfnamefont {D.}~\bibnamefont
  {Bonfiglio}}, \bibinfo {author} {\bibfnamefont {S.}~\bibnamefont {Cappello}},
  \ and\ \bibinfo {author} {\bibfnamefont {D.~F.}\ \bibnamefont {Escande}},\
  }\bibfield  {title} {\enquote {\bibinfo {title} {{Dominant electrostatic
  nature of the reversed field pinch dynamo}},}\ }\href {\doibase
  10.1103/PhysRevLett.94.145001} {\bibfield  {journal} {\bibinfo  {journal}
  {Physical Review Letters}\ }\textbf {\bibinfo {volume} {94}},\ \bibinfo
  {pages} {145001} (\bibinfo {year} {2005})}\BibitemShut {NoStop}%
\bibitem [{\citenamefont {Ji}\ \emph {et~al.}(1996)\citenamefont {Ji},
  \citenamefont {Prager}, \citenamefont {Almagri}, \citenamefont {Sarff},
  \citenamefont {Yagi}, \citenamefont {Hirano}, \citenamefont {Hattori},\ and\
  \citenamefont {Toyama}}]{Ji1996}%
  \BibitemOpen
  \bibfield  {author} {\bibinfo {author} {\bibfnamefont {H.}~\bibnamefont
  {Ji}}, \bibinfo {author} {\bibfnamefont {S.~C.}\ \bibnamefont {Prager}},
  \bibinfo {author} {\bibfnamefont {A.~F.}\ \bibnamefont {Almagri}}, \bibinfo
  {author} {\bibfnamefont {J.~S.}\ \bibnamefont {Sarff}}, \bibinfo {author}
  {\bibfnamefont {Y.}~\bibnamefont {Yagi}}, \bibinfo {author} {\bibfnamefont
  {Y.}~\bibnamefont {Hirano}}, \bibinfo {author} {\bibfnamefont
  {K.}~\bibnamefont {Hattori}}, \ and\ \bibinfo {author} {\bibfnamefont
  {H.}~\bibnamefont {Toyama}},\ }\bibfield  {title} {\enquote {\bibinfo {title}
  {{Measurement of the dynamo effect in a plasma}},}\ }\href {\doibase
  10.1063/1.871989} {\bibfield  {journal} {\bibinfo  {journal} {Physics of
  Plasmas}\ }\textbf {\bibinfo {volume} {3}},\ \bibinfo {pages} {1935--1942}
  (\bibinfo {year} {1996})}\BibitemShut {NoStop}%
\bibitem [{\citenamefont {{Den Hartog}}\ \emph {et~al.}(1999)\citenamefont
  {{Den Hartog}}, \citenamefont {Chapman}, \citenamefont {Craig}, \citenamefont
  {Fiksel}, \citenamefont {Fontana}, \citenamefont {Prager},\ and\
  \citenamefont {Sarff}}]{DenHartog1999}%
  \BibitemOpen
  \bibfield  {author} {\bibinfo {author} {\bibfnamefont {D.~J.}\ \bibnamefont
  {{Den Hartog}}}, \bibinfo {author} {\bibfnamefont {J.~T.}\ \bibnamefont
  {Chapman}}, \bibinfo {author} {\bibfnamefont {D.}~\bibnamefont {Craig}},
  \bibinfo {author} {\bibfnamefont {G.}~\bibnamefont {Fiksel}}, \bibinfo
  {author} {\bibfnamefont {P.~W.}\ \bibnamefont {Fontana}}, \bibinfo {author}
  {\bibfnamefont {S.~C.}\ \bibnamefont {Prager}}, \ and\ \bibinfo {author}
  {\bibfnamefont {J.~S.}\ \bibnamefont {Sarff}},\ }\bibfield  {title} {\enquote
  {\bibinfo {title} {{Measurement of core velocity fluctuations and the dynamo
  in a reversed-field pinch}},}\ }\href {\doibase 10.1063/1.873439} {\bibfield
  {journal} {\bibinfo  {journal} {Physics of Plasmas}\ }\textbf {\bibinfo
  {volume} {6}},\ \bibinfo {pages} {1813--1821} (\bibinfo {year}
  {1999})}\BibitemShut {NoStop}%
\bibitem [{\citenamefont {Fontana}\ \emph {et~al.}(2000)\citenamefont
  {Fontana}, \citenamefont {{Den Hartog}}, \citenamefont {Fiksel},\ and\
  \citenamefont {Prager}}]{Fontana2000}%
  \BibitemOpen
  \bibfield  {author} {\bibinfo {author} {\bibfnamefont {P.~W.}\ \bibnamefont
  {Fontana}}, \bibinfo {author} {\bibfnamefont {D.~J.}\ \bibnamefont {{Den
  Hartog}}}, \bibinfo {author} {\bibfnamefont {G.}~\bibnamefont {Fiksel}}, \
  and\ \bibinfo {author} {\bibfnamefont {S.~C.}\ \bibnamefont {Prager}},\
  }\href {\doibase 10.1103/PhysRevLett.85.566} {\enquote {\bibinfo {title}
  {{Spectroscopic observation of fluctuation-induced dynamo in the edge of the
  reversed-field pinch}},}\ }\bibinfo {type} {Tech. Rep.}\ \bibinfo {number}
  {3}\ (\bibinfo {year} {2000})\BibitemShut {NoStop}%
\bibitem [{\citenamefont {Brower}\ \emph {et~al.}(2002)\citenamefont {Brower},
  \citenamefont {Ding}, \citenamefont {Terry}, \citenamefont {Anderson},
  \citenamefont {Biewer}, \citenamefont {Chapman}, \citenamefont {Craig},
  \citenamefont {Forest}, \citenamefont {Prager},\ and\ \citenamefont
  {Sarff}}]{Brower2002}%
  \BibitemOpen
  \bibfield  {author} {\bibinfo {author} {\bibfnamefont {D.~L.}\ \bibnamefont
  {Brower}}, \bibinfo {author} {\bibfnamefont {W.~X.}\ \bibnamefont {Ding}},
  \bibinfo {author} {\bibfnamefont {S.~D.}\ \bibnamefont {Terry}}, \bibinfo
  {author} {\bibfnamefont {J.~K.}\ \bibnamefont {Anderson}}, \bibinfo {author}
  {\bibfnamefont {T.~M.}\ \bibnamefont {Biewer}}, \bibinfo {author}
  {\bibfnamefont {B.~E.}\ \bibnamefont {Chapman}}, \bibinfo {author}
  {\bibfnamefont {D.}~\bibnamefont {Craig}}, \bibinfo {author} {\bibfnamefont
  {C.~B.}\ \bibnamefont {Forest}}, \bibinfo {author} {\bibfnamefont {S.~C.}\
  \bibnamefont {Prager}}, \ and\ \bibinfo {author} {\bibfnamefont {J.~S.}\
  \bibnamefont {Sarff}},\ }\bibfield  {title} {\enquote {\bibinfo {title}
  {{Measurement of the Current-Density Profile and Plasma Dynamics in the
  Reversed-Field Pinch}},}\ }\href {\doibase 10.1103/PhysRevLett.88.185005}
  {\bibfield  {journal} {\bibinfo  {journal} {Physical Review Letters}\
  }\textbf {\bibinfo {volume} {88}},\ \bibinfo {pages} {4} (\bibinfo {year}
  {2002})}\BibitemShut {NoStop}%
\bibitem [{\citenamefont {Seehafer}(1996)}]{Seehafer1996}%
  \BibitemOpen
  \bibfield  {author} {\bibinfo {author} {\bibfnamefont {N.}~\bibnamefont
  {Seehafer}},\ }\bibfield  {title} {\enquote {\bibinfo {title} {{Nature of the
  $\alpha$ effect in magnetohydrodynamics}},}\ }\href {\doibase
  10.1103/PhysRevE.53.1283} {\bibfield  {journal} {\bibinfo  {journal}
  {Physical Review E - Statistical Physics, Plasmas, Fluids, and Related
  Interdisciplinary Topics}\ }\textbf {\bibinfo {volume} {53}},\ \bibinfo
  {pages} {1283--1286} (\bibinfo {year} {1996})}\BibitemShut {NoStop}%
\bibitem [{\citenamefont {Ji}\ \emph {et~al.}(1995)\citenamefont {Ji},
  \citenamefont {Yagi}, \citenamefont {Hattori}, \citenamefont {Almagri},
  \citenamefont {Prager}, \citenamefont {Hirano}, \citenamefont {Sarff},
  \citenamefont {Shimada}, \citenamefont {Maejima},\ and\ \citenamefont
  {Hayase}}]{Ji1995}%
  \BibitemOpen
  \bibfield  {author} {\bibinfo {author} {\bibfnamefont {H.}~\bibnamefont
  {Ji}}, \bibinfo {author} {\bibfnamefont {Y.}~\bibnamefont {Yagi}}, \bibinfo
  {author} {\bibfnamefont {K.}~\bibnamefont {Hattori}}, \bibinfo {author}
  {\bibfnamefont {A.~F.}\ \bibnamefont {Almagri}}, \bibinfo {author}
  {\bibfnamefont {S.~C.}\ \bibnamefont {Prager}}, \bibinfo {author}
  {\bibfnamefont {Y.}~\bibnamefont {Hirano}}, \bibinfo {author} {\bibfnamefont
  {J.~S.}\ \bibnamefont {Sarff}}, \bibinfo {author} {\bibfnamefont
  {T.}~\bibnamefont {Shimada}}, \bibinfo {author} {\bibfnamefont
  {Y.}~\bibnamefont {Maejima}}, \ and\ \bibinfo {author} {\bibfnamefont
  {K.}~\bibnamefont {Hayase}},\ }\bibfield  {title} {\enquote {\bibinfo {title}
  {{Effect of collisionality and diamagnetism on the plasma dynamo}},}\ }\href
  {\doibase 10.1103/PhysRevLett.75.1086} {\bibfield  {journal} {\bibinfo
  {journal} {Physical Review Letters}\ }\textbf {\bibinfo {volume} {75}},\
  \bibinfo {pages} {1086--1089} (\bibinfo {year} {1995})}\BibitemShut {NoStop}%
\bibitem [{\citenamefont {Furth}, \citenamefont {Killeen},\ and\ \citenamefont
  {Rosenbluth}(1963)}]{Furth1963}%
  \BibitemOpen
  \bibfield  {author} {\bibinfo {author} {\bibfnamefont {H.~P.}\ \bibnamefont
  {Furth}}, \bibinfo {author} {\bibfnamefont {J.}~\bibnamefont {Killeen}}, \
  and\ \bibinfo {author} {\bibfnamefont {M.~N.}\ \bibnamefont {Rosenbluth}},\
  }\bibfield  {title} {\enquote {\bibinfo {title} {{Finite-resistivity
  instabilities of a sheet pinch}},}\ }\href {\doibase 10.1063/1.1706761}
  {\bibfield  {journal} {\bibinfo  {journal} {Physics of Fluids}\ }\textbf
  {\bibinfo {volume} {6}},\ \bibinfo {pages} {459--484} (\bibinfo {year}
  {1963})}\BibitemShut {NoStop}%
\bibitem [{\citenamefont {Rutherford}(1973)}]{Rutherford1973}%
  \BibitemOpen
  \bibfield  {author} {\bibinfo {author} {\bibfnamefont {P.~H.}\ \bibnamefont
  {Rutherford}},\ }\bibfield  {title} {\enquote {\bibinfo {title} {{Nonlinear
  growth of the tearing mode}},}\ }\href {\doibase 10.1063/1.1694232}
  {\bibfield  {journal} {\bibinfo  {journal} {Physics of Fluids}\ }\textbf
  {\bibinfo {volume} {16}},\ \bibinfo {pages} {1903--1908} (\bibinfo {year}
  {1973})}\BibitemShut {NoStop}%
\bibitem [{\citenamefont {Li}(1995)}]{Li1995}%
  \BibitemOpen
  \bibfield  {author} {\bibinfo {author} {\bibfnamefont {D.}~\bibnamefont
  {Li}},\ }\bibfield  {title} {\enquote {\bibinfo {title} {{A new algebraic
  growth of nonlinear tearing mode}},}\ }\href {\doibase 10.1063/1.871162}
  {\bibfield  {journal} {\bibinfo  {journal} {Physics of Plasmas}\ }\textbf
  {\bibinfo {volume} {2}},\ \bibinfo {pages} {3275--3281} (\bibinfo {year}
  {1995})}\BibitemShut {NoStop}%
\bibitem [{\citenamefont {Wesson}(2011)}]{Wesson2011}%
  \BibitemOpen
  \bibfield  {author} {\bibinfo {author} {\bibfnamefont {J.}~\bibnamefont
  {Wesson}},\ }\href@noop {} {\emph {\bibinfo {title} {{Tokamaks}}}},\ \bibinfo
  {edition} {fourth edition}\ ed.\ (\bibinfo  {publisher} {Oxford University
  Press},\ \bibinfo {year} {2011})\ p.\ \bibinfo {pages} {322}\BibitemShut
  {NoStop}%
\bibitem [{\citenamefont {Abramowitz}\ and\ \citenamefont
  {Stegun}(1972)}]{abramowitz1972handbook}%
  \BibitemOpen
  \bibfield  {author} {\bibinfo {author} {\bibfnamefont {M.}~\bibnamefont
  {Abramowitz}}\ and\ \bibinfo {author} {\bibfnamefont {I.}~\bibnamefont
  {Stegun}},\ }\href {https://books.google.com/books?id=Cxsty7Np9sUC} {\emph
  {\bibinfo {title} {Handbook of Mathematical Functions: With Formulas, Graphs,
  and Mathematical Tables}}},\ Applied mathematics series\ (\bibinfo
  {publisher} {U.S. Department of Commerce, National Bureau of Standards},\
  \bibinfo {year} {1972})\ p.\ \bibinfo {pages} {496}\BibitemShut {NoStop}%
\end{thebibliography}%

\end{document}